# Direct observation of a magnetic field-induced Wigner crystal

Yen-Chen Tsui[1,*], Minhao He[1,*], Yuwen Hu[1,*], Ethan Lake[2], Taige Wang[2,3], Kenji Watanabe[4], Takashi Taniguchi[5], Michael P. Zaletel[2,3], Ali Yazdani[1,†]

[1] *Joseph Henry Laboratories and Department of Physics, Princeton University, Princeton, NJ 08544, USA*
[2] *Department of Physics, University of California, Berkeley, Berkeley, CA 94720, USA.*
[3] *Materials Sciences Division, Lawrence Berkeley National Laboratory, Berkeley, CA 94720, USA.*
[4] *Research Center for Functional Materials, National Institute for Materials Science, 1-1 Namiki, Tsukuba 305-0044, Japan*
[5] *International Center for Materials Nanoarchitectonics, National Institute for Materials Science, 1-1 Namiki, Tsukuba 305-0044, Japan*

\* These authors contributed equally to this work.
† Corresponding author email: yazdani@princeton.edu

**Eugene Wigner predicted long ago that when the Coulomb interactions between electrons become much stronger than their kinetic energy, electrons crystallize into a closely packed lattice[1]. A variety of two-dimensional systems have shown evidence for Wigner crystals[2–11] ; however, a spontaneously formed classical or quantum Wigner crystal (WC) has never been directly visualized. Neither the identification of the WC symmetry nor direct investigation of its melting has been accomplished.  Here we use high-resolution scanning tunneling microscopy (STM) measurements to directly image a magnetic field-induced electron WC in Bernal-stacked bilayer graphene (BLG), and examine its structural properties as a function of electron density, magnetic field, and temperature. At high fields and the lowest temperature, we observe a triangular lattice electron WC in the lowest Landau Level (LLL) of BLG. The WC possesses the expected lattice constant and is robust in a range of filling factors between $\nu \sim 0.13$ and $\nu \sim 0.38$ except near fillings where it competes with fractional quantum Hall (FQH) states. Increasing the density or temperature results in the melting of the WC into a liquid phase that is isotropic but has a modulated structure characterized by the WC's Bragg wavevector. At low magnetic fields, the WC unexpectedly transitions into an anisotropic stripe phase, which has been commonly anticipated to form in higher LLs. Analysis of individual lattice sites reveals**

**signatures that may be related to the quantum zero-point motion of electrons in the WC lattice.**

In low density two-dimensional (2D) electronic systems, electrons are predicted to form a Wigner crystal (WC) when the ratio of the Coulomb energy to the kinetic energy becomes larger than ~40 [12]. The application of a perpendicular magnetic field strongly suppresses the electron's kinetic energy by forming Landau levels (LL), thereby favoring WC formation at higher electron densities[3], with its melting transition controlled by the LL's filling factor[13,14]. A variety of 2D systems in both zero and high fields have shown evidence for electron crystallization using a wide range of experimental techniques. Most techniques [2,4–11,15–26] provide indirect evidence for WC formation, making it difficult to distinguish crystallization from localization by disorder, and are unable to provide information on the WC's lattice structure and symmetry[27,28]. Organization of up to six electrons in a carbon nanotube in a pattern consistent with a very small one-dimensional WC has been visualized[29]. While most 2D systems studied are inaccessible to imaging techniques, recent studies of transition metal dichalcogenide layers have succeeded in imaging generalized WC in zero magnetic field[30], the structures of which are dictated by a superlattice moiré potential that traps electrons. These crystalline electron phases are different from the WC phases expected to form spontaneously in the absence of any periodic potential and break continuous as opposed to discrete translational symmetry. Here we build on recent advances in scanning tunneling microscopy (STM) studies of ultra-clean graphene-based materials in high magnetic field[31–33] to directly image WC lattice structure, visualize its melting, and examine its competition with stripe ordering and fractional quantum Hall (FQH) states.

**Evidence of the Wigner Crystal**

STM measurements of graphite-gated ultra-clean Bernal-stacked bilayer graphene (BLG) devices have recently provided high-resolution spectroscopic measurements of a rich array of FQH states[33]. These experiments have enabled probing FQH states in idealized conditions away from any defects, as well as fragile FQH states, such as the even-denominator candidate non-Abelian states, which form in ultra-clean BLG devices with remarkably large energy gaps. The presence of sufficiently large defect-free areas in such ultra-clean BLG devices (as imaged in Fig. 1a) makes them ideal for attempting to visualize WCs by spatially mapping the electronic properties of a partially filled LL. In this study, we focus on the filling range $0 < \nu < 1$, where we have previously shown that electrons occupy the orbital state $N = 0$ in the top graphene layer of the BLG device[33]. Figure 1b shows density-dependent scanning tunneling spectroscopy (DD-STS) measurement in this filling range (at 13.95 T) that shows a tunneling Coulomb gap $\Delta_C$ near zero sample bias $V_B$ (for tunneling into a 2D electron gas at finite field[31]), as well as features associated with the FQH states at $\nu = 1/3, 2/5, 3/7, 4/9, 5/11$. To probe the presence of a WC, we image the spatial modulation of the tunneling current $\delta I_{dc}$ at the threshold bias for electron tunneling with $V_B$ set at the edge of the Coulomb gap, while the STM feedback loop is disabled (see Methods). Such maps, an example of which is shown in Fig. 1c for $\nu = 0.317$, reveal a triangular lattice structure in the modulation of $\delta I_{dc}$, with a periodicity of near 30 nm that is consistent with that of an electron WC as we discuss below.

The fast Fourier transforms (FFT) of $\delta I_{dc}$ map for $\nu = 0.317$ (Fig. 1d) shows six first order Bragg peaks (red circles) and many higher order peaks with well-defined $C_6$ symmetry, indicating a well-ordered WC in regions far from defects. The FFT also shows distinct peaks corresponding to a BLG-hBN incommensurate moiré superlattice (black diamonds) at length scales far smaller than that of the WC. Imaging the WC near defects, we see clear signatures that the WC is pinned by individual defects[34] (see SI Fig. S1). When performing STM imaging with electrons tunneling from the tip into the sample ($V_B > 0$), we observe periodic suppression of tunneling current and modulation of threshold bias voltage, suggesting that imaging contrast is related to the Coulomb suppression for electron tunneling near the WC's lattice sites, where electrons are localized (see SI Fig. S2). Reversing the sample bias ($V_B < 0$) at the threshold for tunneling, we observe an approximate reversal of imaging contrast and the enhancement of tunneling current $I_{dc}$ out of the sample correlated with the suppression of tunneling at positive bias (see SI Fig. S3). This behavior supports an interpretation in which the STM tip has a relatively small impact on the WC. A priori, the potential produced by the STM tip, which contains a component in proportion to $V_B$, could locally distort or even drag the WC beneath the tip. However, since reversing $V_B$ (at low-bias) does not qualitatively impact our observations, this suggests the impact of the tip-produced potential is minimal.

Examining the $\delta I_{dc}$ map as function of $\nu$ at the highest magnetic field and lowest temperature in our study (13.95 T, 210 mK electron temperature), we obtain precise measurements of the WC lattice structure that can be compared to theoretical predictions, as well as find evidence for melting of the WC and its competition with FQH states. Figs. 2a-h show a series of $\delta I_{dc}$ maps measured in the exact same area of the sample while increasing $\nu$ (see full data set in SI Fig. S4), which show the evolving spatial electronic structure of the partially filled LL. To analyze this evolution, we plot the corresponding structure factor $S(\vec{q})$ of the $\delta I_{dc}$ maps in Figs. 2i-p, which are the FFT of their autocorrelation maps (see Methods). At low filling factors $\nu \sim 0.1$, the real space maps show distorted structures, with their $S(\vec{q})$ indicating the absence of an ordered WC presumably due to the importance of intrinsic disorder potential at low densities[35]. In contrast, with increasing filling a well-ordered WC emerges showing a corresponding $S(\vec{q})$ with six sharp Bragg peaks, which show the simultaneous breaking of the translational and rotational symmetries, and the Bragg wavevectors disperse with increasing $\nu$ indicating the decreasing lattice constant (Fig. 2j-m). The continuously tuned peaks in $S(\vec{q})$ between $\nu \sim 0.15$ and $\nu \sim 0.31$ rotate between different densities (Fig. 2j-m), indicating that the orientationally ordered WC is not locked to any specific lattice potential (see SI Fig. S5). At the precise fillings corresponding to FQH states, such as $\nu = 1/3$ shown in Fig. 2f, $S(\vec{q})$ appears to be featureless, but more careful examination of the data near these FQH phases reveals that the competition between WC and FQH states involves an intermediate liquid-like phase that has no orientational order (see SI Fig. S6). Increasing the filling, we find that WC remerges in between $\nu = 1/3$ and $2/5$, up to $\nu_{max} \sim 0.38$ (Fig. 2g, 2o), but eventually at larger fillings it melts into a liquid phase (Fig. 2h) with $S(\vec{q})$ that shows a ring-like feature as shown in Fig. 2p (see SI Fig. S4 for more).

To make a quantitative comparison with the predicted lattice constant of the WC, we extract the Bragg vectors $|\vec{q}|$ from the structure factor $S(\vec{q})$ maps (see Methods) and calculate the corresponding triangular lattice constant $a = \frac{4\pi}{\sqrt{3}|\vec{q}|}$ as a function of $\nu$ for data that show both translational and orientational order. As shown in Fig. 2q, the experimental data follows the predicted lattice constant for the WC lattice $a = \sqrt{\frac{4\pi l_B^2}{\sqrt{3}\nu}}$ (the blue curve) over a large range of $\nu$ remarkably well and confirms that we are imaging a well-ordered WC. Such excellent agreement rules out other phases, such as bubble phase[36–40], and also demonstrates that our measurement conditions are minimally perturbing the WC's structure. We can also extend the Fourier analysis to the liquid phases by extracting the characteristic $|\vec{q}|$ from the structure factor $S(\vec{q})$ maps and characterizing the modulations we observed in our $\delta I_{dc}$ maps. Remarkably, the liquid phases

near FQH phases, such as 1/3 (see SI Fig. S7) and those at higher fillings (Fig. S8), have a ring-like $S(\vec{q})$ that are characterized by a $|\vec{q}|$ very close to those for the WC at the same fillings.

The nature of quantum phase transition between WC and liquid phases such as FQH has been subject of numerous theoretical studies. Possibility of a first-order phase transition, its absence in the presence of disorder, the importance of the long-range Coulomb repulsion forbidding phase separation and prediction of the micro-emulsion intermediate phase, and crystallization of the fractionalized quasi-particles have all been considered [41–45]. In the micro-emulsion scenario, which is relevant to singly-gated devices such as ours, domains of the competing phases interleave at a scale set by the gate distance (55 nm). Our observation of an intermediate liquid phase that appears to be uniform far from defects on length scales of 300 nm (Fig. S4, S6 labeled as green circles) seems to be inconsistent with the possibility of modulated domains of WC and FQH. Examination on longer length scale in samples even cleaner than those studied here would be required to rule out this possibility. Our data does show some signature of inhomogeneous behavior near the phase transition (see Fig. S6, labeled as yellow stars $\nu$ = 0.33-0.332 and 0.335-0.342, where FQH and the intermediate liquid phase appear to coexist); however, they may be associated with defects rather than spontaneous phase separation or micro-emulsion.

The nature of our intermediate correlated liquid phases remains to be fully understood; however, they are clearly distinct from that of a Fermi liquid or composite Fermi liquid[46], which in the presence of defects is expected to have modulated electronic structure associated with scattering on the (spin-polarized) Fermi surface at wavevector $2k_\mathrm{F} = 2\sqrt{4\pi n} = 2\sqrt{2\nu/l_B^2}$, where $k_\mathrm{F}$ is the Fermi wave vector at zero field and $n$ is the electron density (see SI Fig. S8 for comparison). One possibility is that a FQH state with dilute trapped quasi-particles may have liquid-like response for magneto-rotons at a $|\vec{q}|$, which have a dispersion minima at a $|\vec{q}|$ very close to that of the WC[47]. An unknown intermediate phase is also possible; for example, at low magnetic fields Monte Carlo simulations of WC melting have revealed the possibility of an intermediate "hybrid" liquid-like phase[48,49]. We add that the re-entrant behavior of the WC between 1/3 and 2/5 FQH state is similar to that previously reported from screening efficiency measurements on dilute 2D hole system in GaAs/AlGaAs heterostructure[50]; however, the effective mass for this system is an order magnitude larger than BLG[51], which is thought to be responsible for its stability. One possible explanation of the stability of WC in BLG over a comparably wide filling range is the LL mixing between the orbital number $N = 0, 1$ states, which might suppress nearby competing FQH states in favor of the WC phase.

**Temperature, magnetic field dependence and observation of the stripe phase**

Having directly visualized the WC with the expected lattice structure and symmetry at high magnetic field at our lowest temperature and its melting into correlated liquid phases as a function of $\nu$, we turn our attention to its stability at lower fields and higher temperatures. While we have not performed detailed temperature dependent studies, the WC phase is observed to have melted at 3 K, with $\delta I_{dc}$ measured at 13 T at this temperature showing liquid-like $S(\vec{q})$ (Fig. 3a-f, see SI Fig. S9 for more). Lowering the magnetic field at the lowest temperatures (Fig. 3h-u), we find that the WC becomes distorted, with $S(\vec{q})$ not showing all the expected six peaks of a well-ordered triangular lattice. At some fields and densities, the WC transforms into an unexpected stripe phase[52–56] (Fig. 3l and 3s) with an orientation that is not correlated with any structural features of our sample, and which also rotates when we vary field or density (see SI Fig. S10). Figure 3g shows a detailed analysis of the field dependence of $\delta I_{dc}$ maps for different fillings, annotating the different types of $S(\vec{q})$ features characterizing the phases (WC, distorted WC, stripe, and liquid phase) that are observed in each measurement, as well as their characteristic modulation wavevectors at each filling and magnetic field (see Methods and Table S1 for phase characterization and SI Fig. S4, S11-S16 for full data set). At low magnetic fields, the stability of the stripe phase is likely due to stronger LL mixing, which softens the Coulomb interaction as higher LL mixed in. Remarkably, both the stripe and liquid phases we have observed (at both low and high temperatures) are all characterized by a spatial modulation corresponding to a $|\vec{q}|$ very close to that expected for a WC, at the same filling and magnetic fields. The strong correlations creating the WC clearly still determine the structure of the correlated liquid and the stripe phases.

**The quantum nature of the Wigner Crystal**

Finally, we study the quantum and thermal fluctuations of the WC by investigating the spatial structure of individual WC sites. Of particular interest is the variance $\sigma$ of the tunneling current profile at each WC site, which we associate with the spatial extent of electron wavefunctions in the WC. An extraction of $\sigma$ (see Methods and SI Fig. S17) yields the results presented in Fig. 4a, which show that as $\nu$ is increased, $\sigma$ exhibits a pronounced decrease, followed by a saturation approximately around $\sqrt{2}\, l_B$ before melting. Remarkably, we observe the ratio $\sigma/a$ to be rather large (~ 0.3) and roughly independent of $\nu$ over a wide range of fillings and fields (Fig. 4b). The importance of quantum zero-point motion is typically captured using a quantity similar to our $\sigma/a$, the so-called de Boer parameter, which is the ratio of de Broglie wavelength to the lattice constant. The extreme example of a quantum crystal[57] is solid $^4$He, with a de Boer parameter close to 0.4, which is larger but close to the value of $\sigma/a \sim 0.3$ for our WC. We also note that $\sigma$ remains roughly unchanged even when the WC distorts and begins to melt at large $\nu$, suggesting that the mechanism of melting is distinct from the usual Lindemann theory[58,59].

A minimal theoretical model addressing these observations is one in which the variance of electron positions is assumed to be caused by the phonon modes of the resulting WC. We have performed such calculations (see SI section 1.1) and find that the simplest model predicts a $\sigma$ about 30-50% lower than observation. The underestimation of $\sigma$ suggests a smaller restoring force in the phonon model, which can be attributed to softening of the Coulomb interaction due to screening and/or mixing of the $N = 0, 1$ LLs. A phonon model with RPA screening predicts $\sigma$'s that are closer to the experimental values at high $\nu$ (see SI section 1.1) but deviates at low fillings. Such discrepancy and larger $\sigma$ in general could also be due to the tip-sample interaction, which we expect to be stronger at lower $\nu$, where the WC is softer and hence more easily to be distorted by the presence of the tip (see SI section 1.2). Disentangling the role of softened Coulomb interaction from tip perturbations requires further systematic measurements as function of temperature and tunneling bias $V_B$.

**Discussion**

Looking ahead, our imaging technique can potentially be used to examine a wide range of spatially modulated electronic phases, including hole WC, WC at zero magnetic field, bilayer WC, Skyrme crystal, bubble phases, and possibly WC of quasiparticles found near fractional quantum Hall states. Detailed structural studies could also be used to examine the mechanism of melting such as topological defect mediated melting mechanism[60,61]. Measurements in the presence of a net electrical current flowing through the device could also be used to explore the depinning of a WC from the disorder potential.

## Methods

**Sample Preparation**

The BLG devices are fabricated with a modified dry transfer method. The heterostructure consists of exfoliated BLG/hBN/few layer graphite layers from top to bottom. The heterostructure is picked up by a polyvinyl alcohol (PVA) handle, supported by transparent tape on polydimethylsiloxane (PDMS), and then transferred onto a pre-patterned $SiO_2$ (285 nm)/Si substrate with Au/Ti contacts (50 nm/3 nm). We focus on a device with the hBN dielectric layer of 55 nm in this study. The PVA film is first dissolved by HPLC water. Then the sample surface is extensively cleaned in a series of solvent including HPLC water, acetone, isopropyl alcohol (IPA), and n-methyl-2-pyrrolidone (NMP). The device is then transferred into the ultra-high vacuum chamber to anneal at 475 °C overnight before transferring into STM.

**STM Measurement**

STM experiment is performed in a home-built dilution refrigerator STM with mixing chamber temperature ~20 mK and effective electron temperature $T_{eff}$ ~ 210 mK calibrated on a single crystal Al(100) unless specified. Data in this paper is obtained with a perpendicular magnetic field ranging from 2 T to 13.95 T. The measurements are performed by tungsten tips carefully prepared on Cu(111) surface to minimize the work function mismatch between the tip and the sample[32].

The measurements are performed with tip grounded and a bias voltage $V_B$ applied to the BLG. A total of $V_B+V_G$ is applied to the graphite back gate to maintain a relative $V_G$ difference between the sample and the back gate. The differential conductance $dI/dV$ is taken by lock-in method with AC modulation at a frequency of 712.9 Hz in typical STS measurements.

The large area tunneling current map ($\delta I_{dc}$ map) is taken with the following procedure: For each line of the map, the first forward/backward scan is performed with a setpoint $V_B$ = -0.2 V, $I$ = 1 nA at constant current mode to record the topography corrugation z (the typical scan speed is ~3 seconds per line). For the second forward/backward scan, the tip returns to its starting position and then follows the recorded forward/backward height z (with feedback disabled) with a small bias voltage $V_B$ close to the Fermi energy and record the DC current channel $I_{dc}$ at the same time. Lastly, the tip returns to its starting position of each line, moves upwards to the next pixel and then repeats the above line scan until the map is done. To capture the underlying electronic structure clearly, we present the spatial modulation of the dc tunneling current $\delta I_{dc}$ by subtracting its line average: $\delta I_{dc} = |I_{dc}| - \langle |I_{dc}| \rangle_{line}$. Notice that for $V_B > 0$, $\delta I_{dc} = I_{dc} - \langle I_{dc} \rangle_{line}$ since the current is always positive; while for $V_B < 0$, $\delta I_{dc} = -(I_{dc} - \langle I_{dc} \rangle_{line})$ because the current is negative. Therefore, large $\delta I_{dc}$ means the enhancement of the current signal, while small $\delta I_{dc}$ means the suppression of the current. Note that for $V_B > 0$ ($V_B < 0$), positive (negative) DC current $I_{dc}$ is always obtained while its spatial modulation $\delta I_{dc}$ could be a negative (positive) value as indicated in the color bar.

**Structure factor $S(\vec{q})$ extraction**

The structure factor of the $\delta I_{dc}$ map is defined as $S(\vec{q}) \equiv \langle \delta I_{dc}(\vec{q}) \times \delta I_{dc}(-\vec{q}) \rangle$ where $\delta I_{dc}(\vec{q}) = \int \delta I_{dc}(\vec{r}) e^{i\vec{q}\cdot\vec{r}} d\vec{r}$. Mathematically, the Fourier Transform (FT) of an autocorrelation function (ACF) of the $\delta I_{dc}$ map, which is the power spectral density (PSD), is equivalent to the modulus squared Fast Fourier Transform (FFT) map of the $\delta I_{dc}$ map ($|\delta I_{dc}(\vec{q})|^2 = S(\vec{q})$). In practice, an autocorrelation function (ACF) map of the $\delta I_{dc}$ is first obtained with zero padding, and the structure factor $S(\vec{q})$ map is then generated by taking a fast Fourier transform (FFT) of the ACF map.

**Determination of the lattice constant $a$**

The lattice constant $a$ is obtained from the reciprocal relation $\vec{a}_i \cdot \vec{q}_j = 2\pi \delta_{ij}$, where we have $a = \frac{2\pi}{|\vec{q}|} \times \frac{2}{\sqrt{3}}$ for the triangular lattice of the WC. The $|\vec{q}|$ is obtained from the structure factor $S(\vec{q})$ map: The $S(\vec{q})$ map is first converted into polar coordinates $(\rho, \theta)$ and then averaged over $\theta$. $|\vec{q}|$ is then extracted by performing a Gaussian fit to the location of the Bragg peak.

**Determination of the electronic ground state**

The electronic ground states are assigned mainly according to the structure factor $S(\vec{q})$ map of measurements at each filling factor and magnetic field, assisted by the corresponding $\delta I_{dc}$ map and its autocorrelation function. Images with six distinguishable Bragg peaks in $S(\vec{q})$ map can be identified with quasi-long-range translational order and quasi-long-range orientational order, and thus are identified as Wigner crystals (WC). Images with distorted, azimuthally broadened, or interconnected six Bragg peaks in $S(\vec{q})$ are assigned as distorted Wigner crystals (dWC) due to lack of the quasi-long-range translational order or orientational order. Images with ring-like features in $S(\vec{q})$ and an isotropic electronic ground state, with the corresponding $\delta I_{dc}$ map showing interference patterns of concentric ring-like structures centered at disorder locations are assigned as liquid states. Images with two main Bragg peaks in $S(\vec{q})$ and strongly anisotropic features in the corresponding $\delta I_{dc}$ map are assigned as stripe phases.

**Extraction of the variance $\sigma$ of an electron site**

The variance $\sigma$ of electron sites is extracted from areas far from impurities, in images identified as WC or distorted WC (see SI Figure S17 for example). The locations of the electron sites are first determined as $(x_i, y_i)$, $i = 1, 2, \ldots, N$, with N being the number of electron sites within the fitting range. The simulated WC lattice consists of electrons on each site with wavefunction of a gaussian form, as expected in the LL with orbital number $N = 0$. The Gaussians are placed on each site with the same variance $\sigma$ and the same amplitude $A$ as fitting parameters, resulting in an electron distribution function:

$$\sum_{i=1}^{N} \sum A e^{-\frac{(x-x_i)^2 + (y-y_i)^2}{2\sigma^2}}$$

This distribution function is then used to fit the normalized tunneling current modulation $\delta I_{dc}$ to find the averaged variance $\sigma$ of electron sites. We interpret $\sigma$ as the size of an electron site in the WC phase.

The variance $\sigma$ fitted above reflect information within a local fitting range, which can vary from region to region depending on the microscopic details such as disorder potential. Measurements at different regions on the sample are performed to extract $\sigma$ following the same procedure. The variance of the $\sigma$, from region to region, is estimated to be ~0.50 nm, which is interpreted as a statistical error of $\sigma$ (distinct from the fitting error as shown in Fig. 4a).

**Acknowledgements**
We acknowledge fruitful discussions with D. Huse, S. Kivelson and M. Heiblum. This work was primarily supported by DOE-BES grant DE-FG02-07ER46419 the Gordon and Betty Moore Foundation's EPiQS initiative grants GBMF9469 to A.Y. Other support for the experimental inferstructure was provided by NSF-MRSEC through the Princeton Center for Complex Materials NSF-DMR- 2011750, ARO MURI (W911NF-21-2-0147), and ONR N00012-21-1-2592. A.Y. acknowledge the hospitality of the Aspen Center for Physics, which is supported by National Science Foundation grant PHY-1607611, where part of this work was carried out. MZ and TW were supported by the U.S. Department of Energy, Office of Science, Office of Basic Energy Sciences, Materials Sciences and Engineering Division, under Contract No. DE-AC02-05CH11231, within the van der Waals Heterostructures Program (KCWF16).

**Author Contributions**
**Y-CT, MH, YH, AY devised the experiments, with Y-CT, MH, YH created devices structures and carried out the STM measurements and data analysis. EL, TW, and MZ carried out the theoretical calculations. KW, and TT provided the h-BN substrates. All authors contributed to writing of the manuscript.**

**Data Availability**
The data that supports the findings of this study are available from the corresponding author upon reasonable request.

**Code Availability**
The code that supports the findings of this study is available from the corresponding author upon reasonable request.

**Competing Interests**
The authors declare no competing interests.

# Figure 1

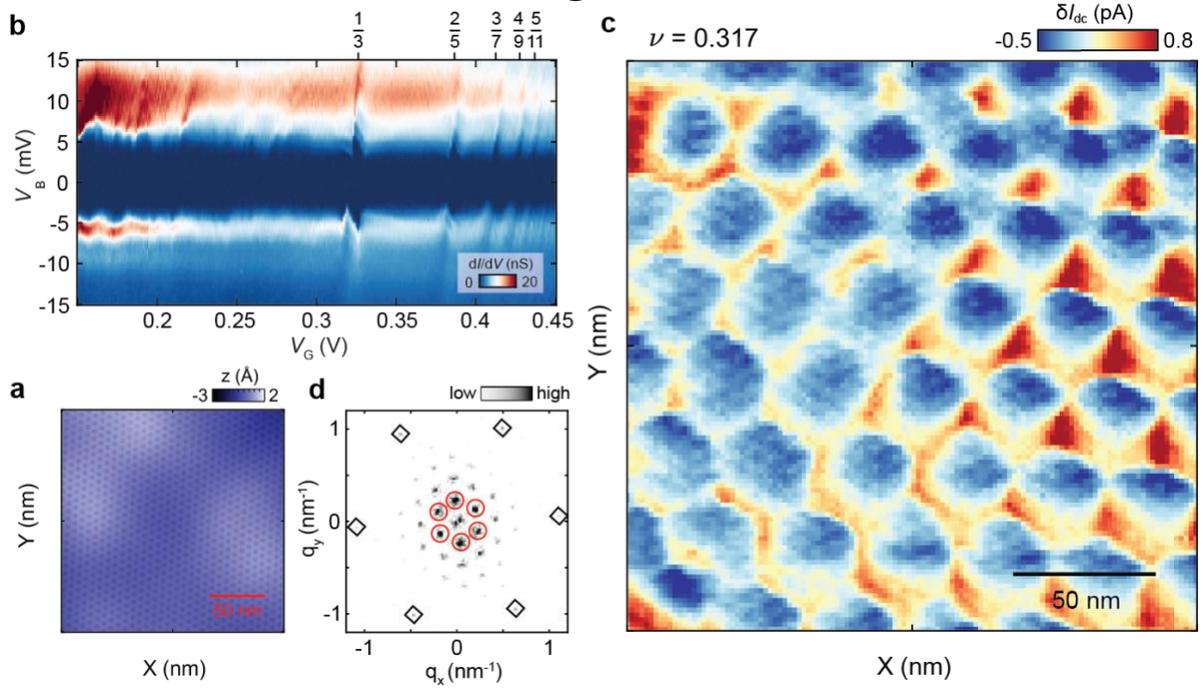

**Figure 1 | Emergent triangular lattice at partial filling of a $N = 0$ Landau level in BLG. a,** A topography image shows clean surface without defects with setpoint $V_B = -0.2$ V, $I = 1$ nA. The periodic structure is the BLG/hBN moiré superlattice. **b,** Density-dependent scanning tunneling spectroscopy (DD-STS) of a partially filled $N = 0$ Landau level in BLG at magnetic field $B = 13.95$ T. The corresponding filling factor $\nu$ of the FQH states are marked on the top. **c,** Spatially resolved tunneling current modulation $\delta I_{dc}$ within a 200 nm by 200 nm region (exact same area as in panel **a**). The measurement is taken with $V_B = 4.6$ mV at filling factor $\nu = 0.317$. **d,** Fast Fourier Transform (FFT) of the tunneling current modulation $\delta I_{dc}$ in panel **c**. The inner peaks in red circles are the first order Bragg peaks of the emergent triangular lattice, while the outer peaks in black diamonds correspond to BLG/hBN moiré lattice with ~ 6.7 nm periodicity.

# Figure 2

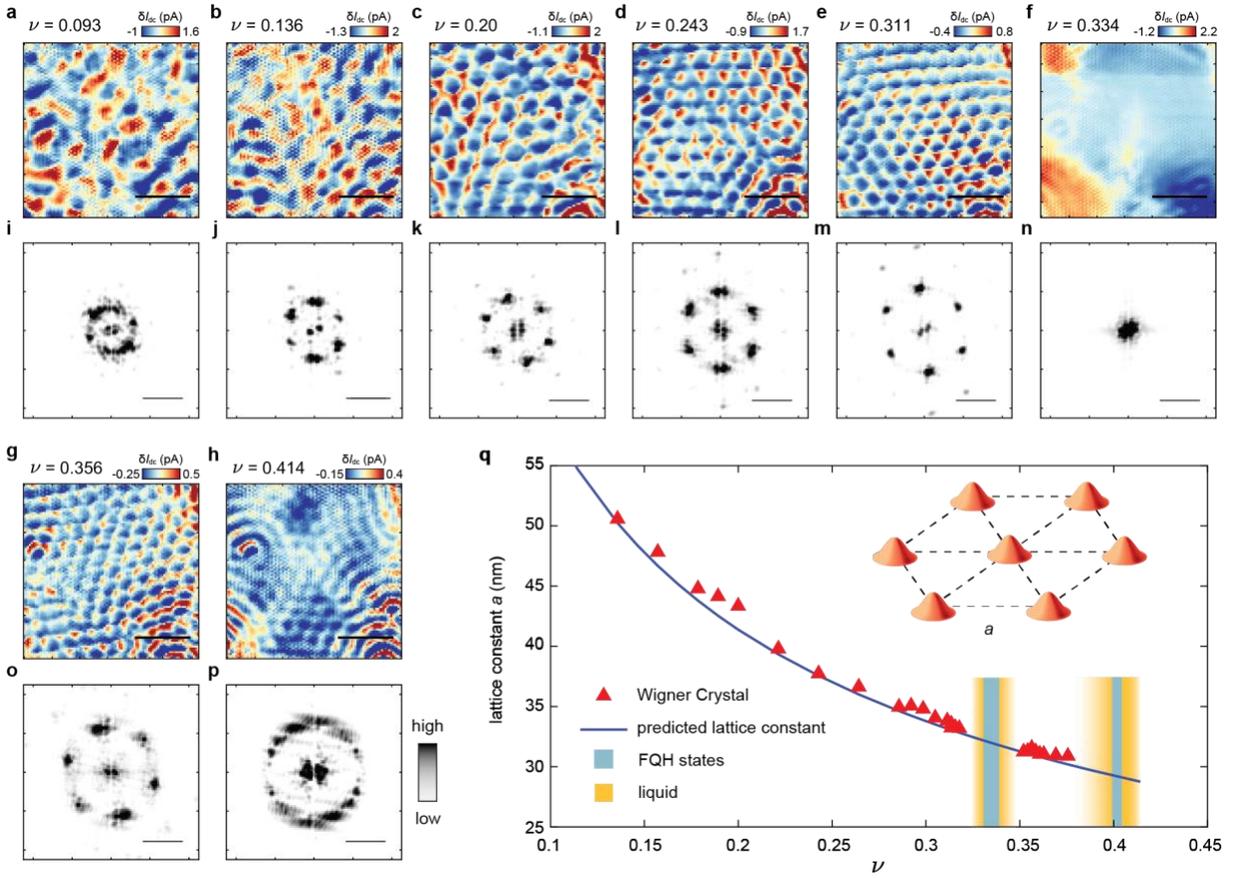

**Figure 2 | Identification of the Wigner Crystal. a-h,** Spatially resolved tunneling current modulation $\delta I_{dc}$ of the same area measured at a series of different filling factor $\nu$, with $V_B$ = 5.2, 5.2, 4.6, 4.4, 4.4, 7.2, 8, 8.8 mV, respectively, at magnetic field $B$ = 13.95 T. The scale bar is 100 nm. **i-p,** Structure factor $S(\vec{q})$ of the tunneling current modulation $\delta I_{dc}$ in panel **a-h** correspondingly. The scale bar is 0.2 nm$^{-1}$. **q,** The extracted lattice constant $a$ of the emergent triangular lattice at different $\nu$ (marked as solid red triangles), measured in the exact same area at $B$ = 13.95 T. The blue line is the expected periodicity of the Wigner Crystal. The red hollow circles are the liquid phase near $\nu$ = 1/3 and at high density. The blue shaded regions mark the incompressible FQH states at $\nu$ = 1/3 and 2/5. The yellow shaded regions mark the liquid phase space. The inset shows a schematic of Wigner crystal, consisting of electrons arranged in a triangular lattice with lattice constant $a$.

# Figure 3

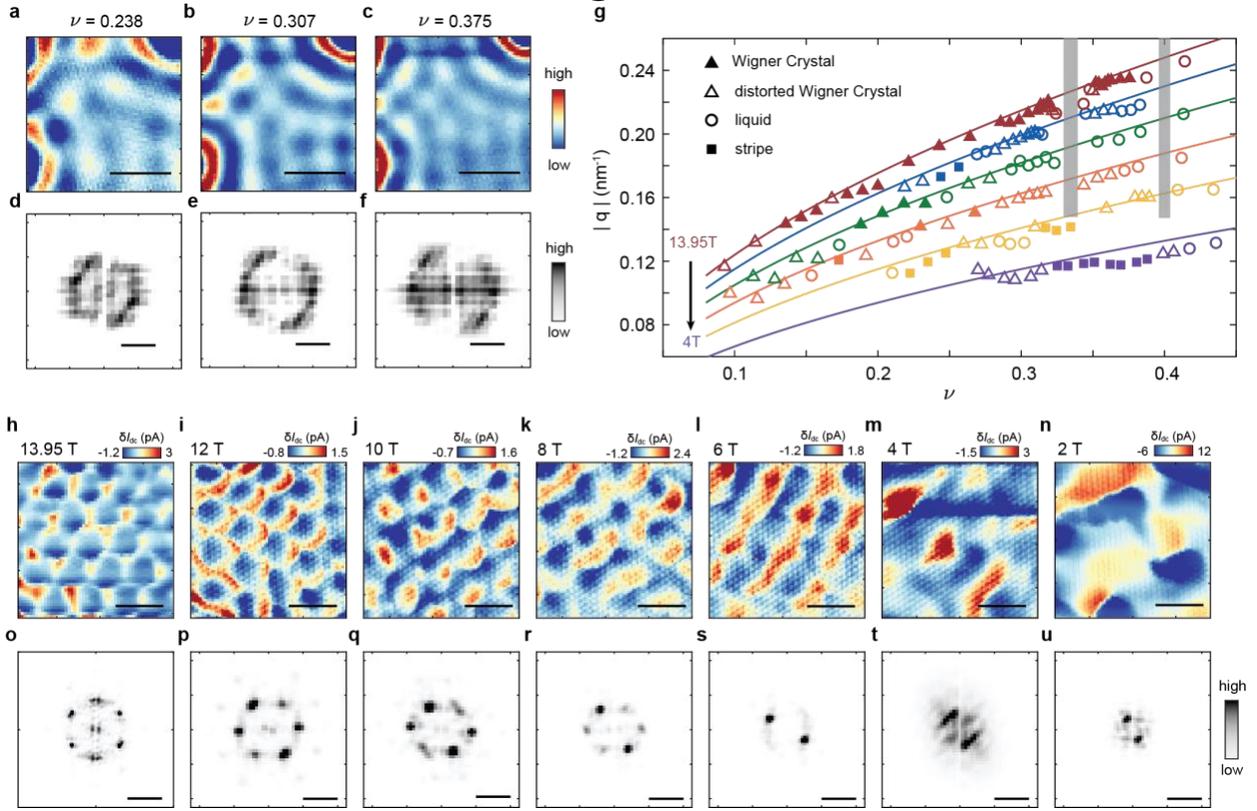

**Figure 3 | Temperature and magnetic field dependence of WC and observation of stripe phase. a-c,** Tunneling current modulation $\delta I_{dc}$ measured at $B$ = 13 T at $T \sim$ 3 K. **d-f,** Structure factor $S(\vec{q})$ of $\delta I_{dc}$ in panel **a-c**, showing ring-like liquid structure with changing periodicity. **g,** Phase diagram of the electron Wigner crystals, distorted Wigner crystals, stripes, and liquid states observed as a function of magnetic field $B$ and filling factor $\nu$. Wigner crystal, distorted Wigner crystal, liquid phase and stripe phase are represented by filled triangle, hollow triangle, hollow circle and filled square symbol, respectively. Each image is represented by $|\vec{q}|$ of the Bragg peaks in the structure factor $S(\vec{q})$. Red, blue, green, orange, yellow and purple represent data taken at $B$ = 13.95 T, 12 T, 10 T, 8 T, 6 T and 4 T respectively, with the solid lines represent the expected $|\vec{q}|$ of the WC at corresponding $B$ field. **h-n,** Spatially resolved tunneling current modulation $\delta I_{dc}$ measured at different magnetic fields $B$ = 13.95 T, 12 T, 10 T, 8 T, 6 T, 4 T, 2 T respectively. With bias voltages $V_B$ = 5.2, 8, 7.8, 7.3, 7, 6, 3 mV respectively. The measurements are performed at filling factor $\nu \sim$ 0.23. The scale bar is 50 nm. **o-u,** Structure factor $S(\vec{q})$ of the tunneling current modulation $\delta I_{dc}$ in panel **h-n** correspondingly. The stripe phase forms at $B$ = 6 T. At low field $B$ = 4 T, 2 T, no periodic structure is observed. The scale bar is 0.2 nm$^{-1}$.

# Figure 4

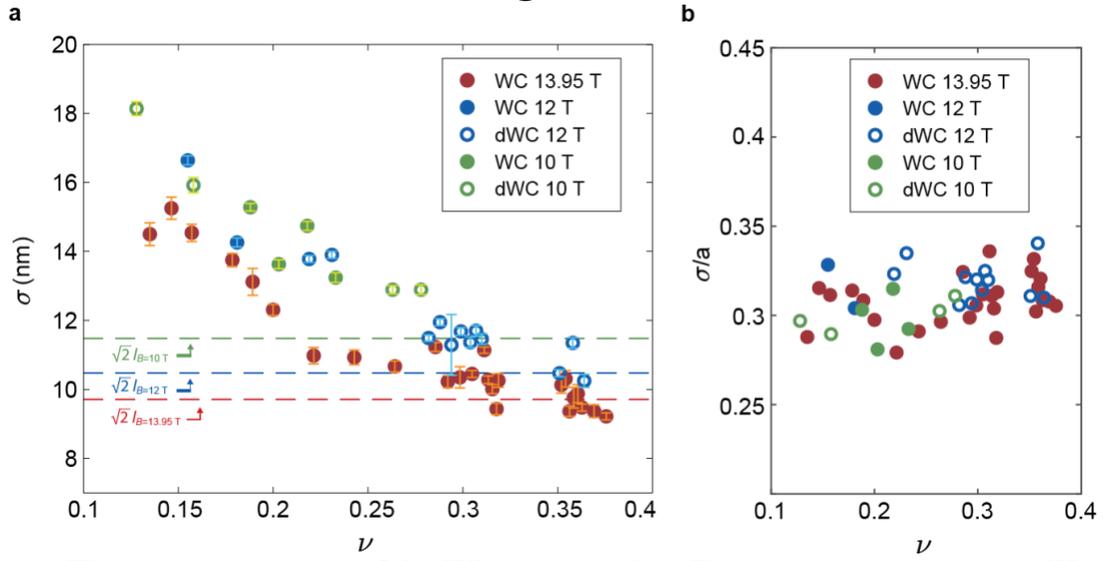

**Figure 4 | The quantum nature of the Wigner crystal. a,** Extracted variance $\sigma$ of the WC sites versus filling factors $\nu$ at $B = 13.95$ T, 12 T, and 10 T, represented by red, blue and green color respectively. The data points show a decreasing trend and saturates at high filling factors. The saturation value scales with magnetic length as around $\sqrt{2}l_B$. The solid circles and hollow circles stand for variance $\sigma$ of the sites extracted from WC and distorted WC respectively. The dashed lines correspond to $\sqrt{2}l_B$ at each field. Error bars represent the fitting error using the Gaussian model (see Methods). $\sigma$ extracted for each field are from the data set measured in the same area. **b,** Ratio of the WC sites variance $\sigma$ and lattice constant $a$, $\sigma/a$ as a function of $\nu$. The ratio remains at a high value ~ 0.3 and is almost constant with respect to $\nu$ and to different fields.